\documentclass{IEEEcsmag}

\usepackage[colorlinks,urlcolor=blue,linkcolor=blue,citecolor=blue]{hyperref}

\usepackage{subcaption}
\usepackage{cite}
\usepackage{upmath}
\usepackage{amsmath}
\usepackage{xcolor}
\usepackage{soul}
\usepackage{booktabs}
\usepackage{pifont}
\newcommand{\xmark}{\ding{55}}

\definecolor{stelios_colour}{RGB}{144, 238, 144}

\newif\ifcomment

\commenttrue

\ifcomment
\newcommand{\stelios}[1]{\sethlcolor{stelios_colour}\hl{[\textbf{Stelios:} #1]}}
\newcommand{\cut}[1]{\sethlcolor{light_red}\hl{[#1]}}
\else
\newcommand{\stelios}[1]{}
\newcommand{\cut}[1]{}

\fi

\jvol{XX}
\jnum{XX}
\paper{8}
\jmonth{May}
\jname{IEEE Computer}
\pubyear{2022}

\usepackage{background}
\backgroundsetup{angle=0,
    scale=1,
    color=black,
    firstpage=true,
    position=current page.north,
    hshift=-20pt,
    vshift=50pt,
    contents={\ifnum\value{page}=1 PREPRINT: Accepted for publication at the IEEE Computer journal, 2022 \else \fi}
}

\begin{document}


\title{Multi-DNN Accelerators for Next-Generation AI Systems}

\author{Stylianos I. Venieris}
\affil{Samsung AI}

\author{Chistos-Savvas Bouganis}
\affil{Imperial College London}

\author{Nicholas D. Lane}
\affil{University of Cambridge\\ Samsung AI}

\markboth{Department Head}{Paper title}

\begin{abstract}
As the use of AI-powered applications widens across multiple domains, so do increase the computational demands. Primary driver of AI technology are the deep neural networks (DNNs).
When focusing either on cloud-based systems that serve multiple AI queries from different users each with their own DNN model, or on mobile robots and smartphones employing pipelines of various models or parallel DNNs for the concurrent processing of multi-modal data, the next generation of AI systems will have multi-DNN workloads at their core.
Large-scale deployment of AI services and integration across mobile and embedded systems require additional breakthroughs in the computer architecture front, with processors that can maintain high performance as the number of DNNs increases while meeting the quality-of-service requirements, giving rise to the topic of multi-DNN accelerator design.
\end{abstract}

\maketitle

\begin{figure*}[t]
    \captionsetup[subfigure]{labelformat=empty}
    \centering
    \vspace{-0.7cm}
    \subfloat[]{
        \centering
        {
        \includegraphics[trim={0.5cm 12.5cm 14.5cm 0.5cm},clip,width=0.33\textwidth]{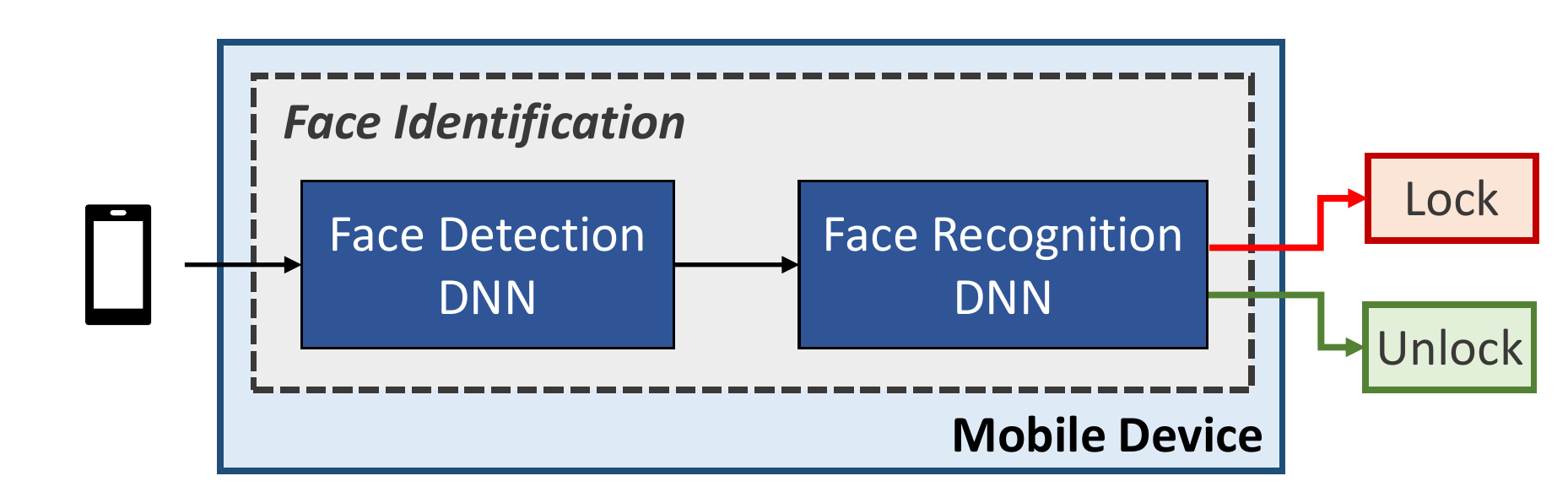}
        }
        \put(-138,2){\scriptsize (a)}
        \label{fig:face-id}
    }
    \subfloat[]{
        \centering
        \includegraphics[trim={0.5cm 7cm 16cm 3cm},clip,width=0.33\textwidth]{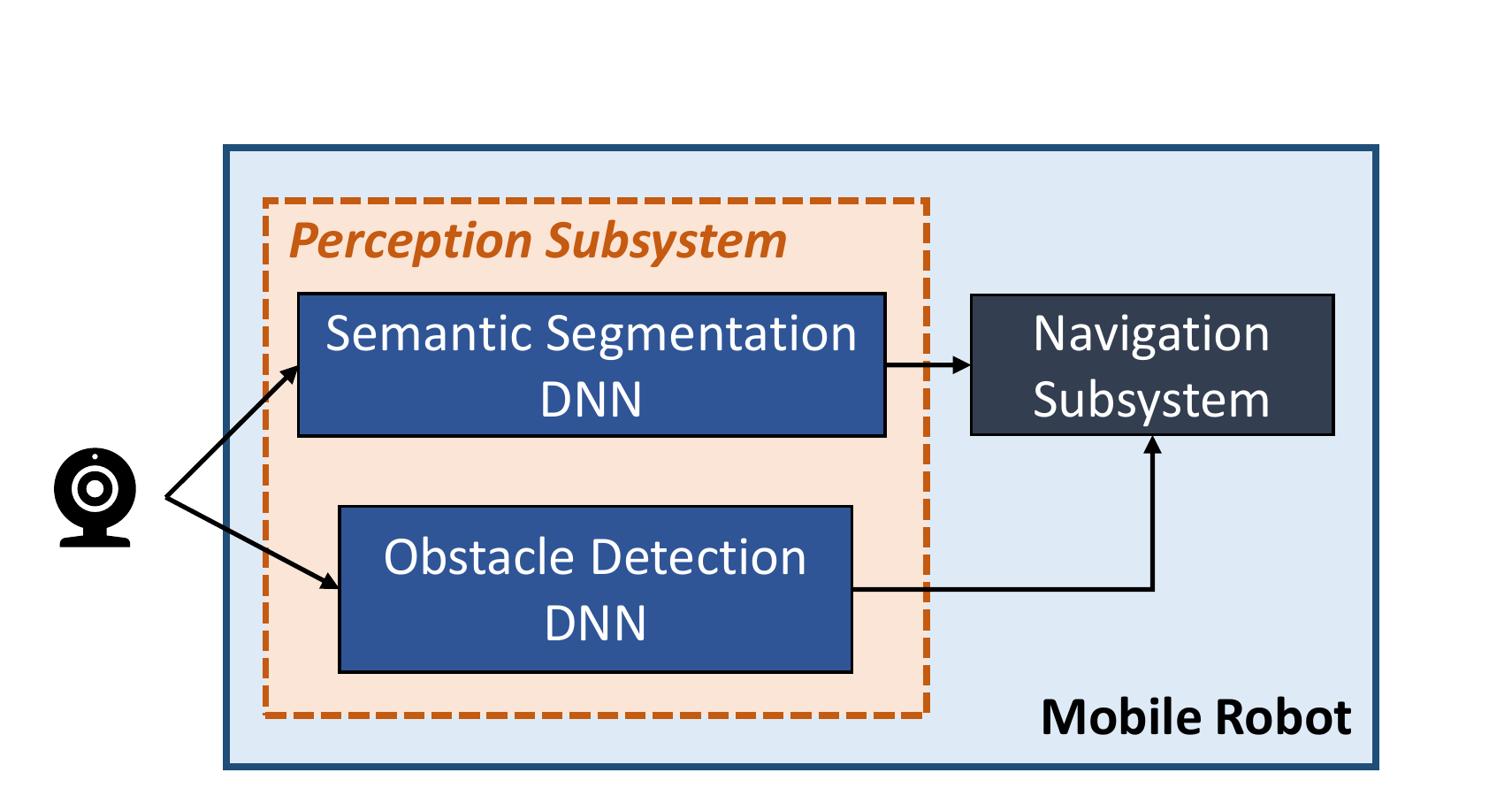}\put(-138,2){\scriptsize (b)}
        \label{fig:mobile-robot}
    }
    \subfloat[]{
        \centering
        \includegraphics[trim={0.5cm 8.2cm 16cm 3cm},clip,width=0.33\textwidth]{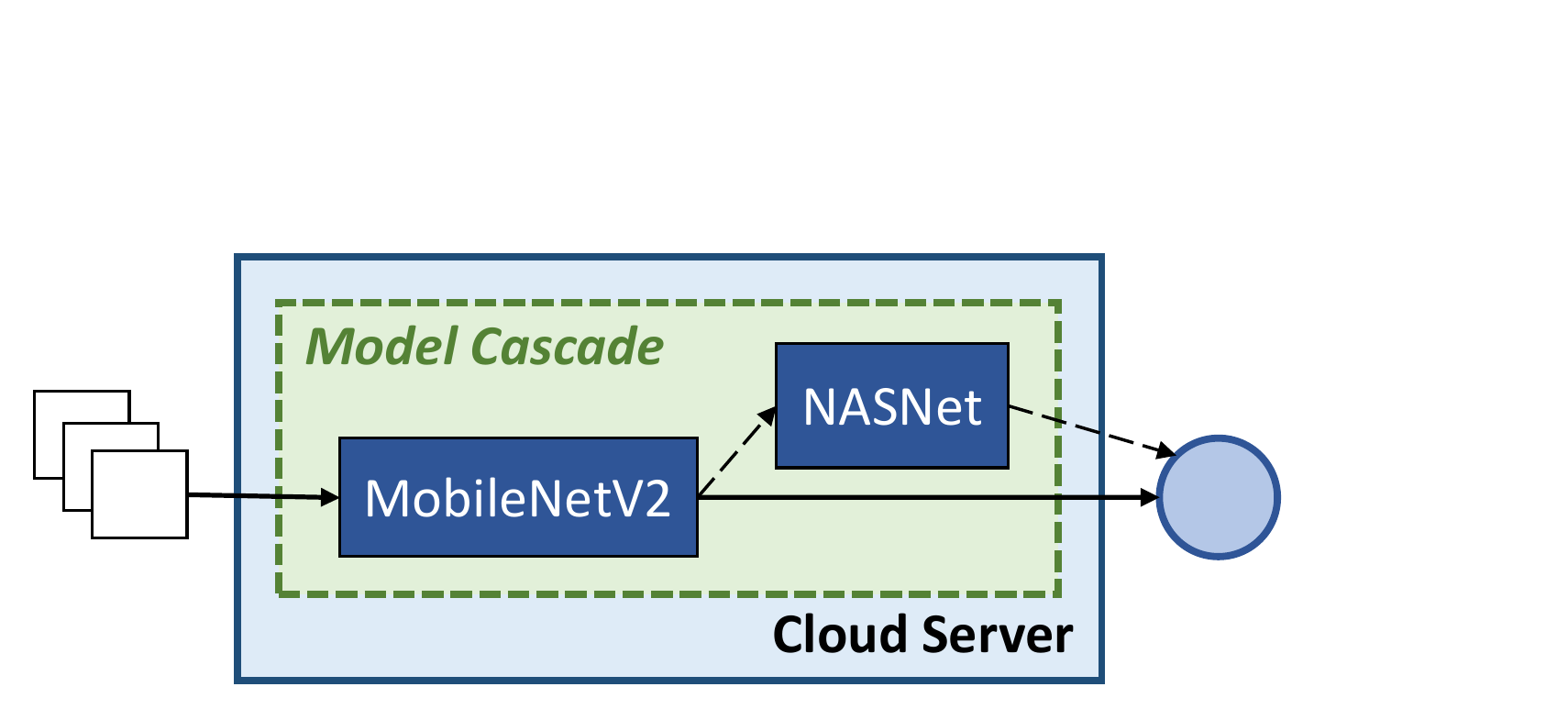}\put(-138,2){\scriptsize (c)}
        \label{fig:model-cascade}
    }
    \vspace{-0.4cm}
    \caption{Multi-DNN systems appear either as grouped DNNs in pipelines (\ref{fig:face-id}), in parallel performing independent tasks (\ref{fig:mobile-robot}) or cascaded for efficient inference (\ref{fig:model-cascade}).}
    \label{fig:multidnn-platforms}
    \vspace{-0.4cm}
\end{figure*}

\chapterinitial{Over the last decade,} deep neural networks (DNNs) have substantially improved the performance of diverse artificial intelligence (AI) tasks. As such, DNNs are seen as the key technology for novel applications in embedded, mobile and cloud setups. One one end, in the embedded space, the emerging field of autonomous robots and vehicles has seen large technological advances based on DNN technology, gathering wide interest due to its potential societal and economic impact, where at the same time, an increasing number of mobile apps are utilizing DNNs for their core functionality. On the other end, cloud-based analytics platforms that employ DNNs are becoming a widespread operational model for serving a large and diverse pool of queries.

Both embedded and cloud AI systems are increasingly integrating \mbox{\textit{multiple DNNs}}. In vision-centric autonomous systems, perception largely relies on highly accurate and reliable computer vision tasks, such as object detection and semantic segmentation. Similarly, smartphones employ pipelines of multiple DNNs in order to improve the quality of the camera-captured content or provide robust augmented reality (AR) functionality.
Cloud-based systems have to cope with servicing a wide range of concurrent DNN workloads, from visual search to speech recognition, with tight real-time constraints. As a result, across all settings, there is a common requirement for the high-performance execution of multiple DNNs.

So far, the computer architecture world has focused on the design of single-DNN accelerators, optimizing metrics crafted for single-DNN execution. With the emergence of multi-DNN applications, there is a need for a new approach in accelerator design that places its focus on the demands and characteristics of such applications. Departing from the single-DNN accelerator paradigm, this new class of multi-DNN computer architectures should process multiple DNNs, capitalizing on properties that are unique to DNN workloads, such as cross-DNN similarities and resilience to reduced precision, and aiming to deliver on multi-DNN metrics, such as maximizing the combined throughput while satisfying the individual latency constraints. 

To this end, there is a need to re-examine several concepts related to DNN accelerator design through the lens of multi-DNN systems, in order to design the next generation of AI accelerators. This process involves the radical rethinking of well-established architectural decisions that were tailored for single-DNN execution, a reprioritization of which hardware components are crucial, and the eventual development of novel and efficient multi-DNN accelerator designs.

However, there are several factors that increase the complexity of designing a multi-DNN accelerator, necessitating further the need for special treatment. First, DNNs come in various forms, giving rise to \mbox{\textit{diverse workloads}}. Depending on the characteristics of the target task, DNNs adopt different topologies, types and number of layers. This directly impacts critical workload dimensions, including the memory footprint, number of operations, computation-to-communication ratio, parallelization potential and resilience to approximate computing techniques. 
Second, many applications consist of \textit{pipelines} of DNNs that feed into each other (\textbf{Figure}~\ref{fig:multidnn-platforms}), creating dependencies and calling for careful architecture design and co-scheduling. Third, depending on the deployment scenario, the demands vary significantly in terms of accuracy, latency and throughput across DNNs, leading to wide \mbox{\textit{performance objectives' variability}} and \textit{multi-objective requirements}.

In this paper, we argue that multi-DNN accelerators can be the key component that drives the next generation of AI systems. We start by presenting their processing requirements and performance criteria. Then, we discuss the current progress in multi-DNN computer architectures and identify their major components, the common challenges and various optimization techniques. Finally, we conclude by outlining open questions and future research directions.

\section{2 OBJECTIVES OF A MULTI-DNN ACCELERATOR}
\label{sec:objectives}

The design of multi-DNN accelerators calls for a shift of methodological paradigm compared to the more mature design flows for conventional single-DNN accelerators. This stems from two orthogonal issues: \textit{i)}~the diversity of workloads in future AI systems and \textit{ii)}~the unique performance criteria of multi-DNN systems.

\subsection{\large\textbf{2.1 Workload Diversity}}
Multi-DNN accelerators are required to support a wide range of AI applications while remaining as future-proof as possible, in order to make efficient use of their resources and amortize their cost. As such, being able to identify and serve future DNN workloads plays a decisive role in the design of a multi-DNN accelerator.

Nevertheless, this is challenged by the current trend towards \textit{task-specialized families of model architectures}. To push the accuracy of each task, the ML community develops DNNs that are optimized for the target task, leading to diverse topologies and, in turn, to workloads with different needs and characteristics. 

For instance, object recognizers and detectors typically employ compute-bound CNNs, while video analyzers are increasingly relying on memory-intensive 3D CNNs.
Other tasks such as image/video super-resolution and semantic segmentation maintain high-frequency details about the input image throughout the model in order to produce high-quality outputs. This property leads to an order of magnitude higher computational and memory demands over classification DNNs.

For natural language processing (NLP) and automatic speech recognition (ASR) tasks, existing systems are dominated by RNNs (\textit{e.g.}~LSTMs/GRUs) and Transformers, with the latter also gaining traction for computer vision tasks. In contrast to the aforementioned DNNs that are primarily compute-bound, these model families are dominated by matrix-vector multiplications and are hence memory-bound. 
As such, designing an accelerator that can serve with high performance both compute- and memory-bound DNNs constitutes a major challenge.

\subsection{Upcoming DNN Workloads}
In addition to the traditional DNN workloads, there is growing interest for new classes of models, with unique characteristics that further diversify the DNN workloads of the future.

\subsubsection{{Dynamic DNNs}}
Recently, numerous adaptive DNN architectures have been proposed~\cite{adaptive_dnns2021emdl}. With the goal of exploiting the variability in complexity across inputs, this class of DNNs can tunably scale their computational needs through either layer/channel skipping or early-exiting mechanisms. Despite their theoretical benefits, dynamic models break the conventionally \textit{static workload} of DNNs - a property that has been broadly exploited to design DNN-tailored accelerators.
\\\textbf{Implications:}
The uncertainty on scheduling poses severe challenges in both designing an efficient pipeline and co-scheduling dynamic models in multi-DNN settings.
As such, there is a need for new hardware techniques that are able to efficiently cope with dynamicity.

\setlength{\tabcolsep}{2pt}
\begin{table*}[t]
    \centering
    \caption{Comparison of Multi-DNN Accelerators}
    \resizebox{1\linewidth}{!}{
    \scriptsize
    \begin{tabular}{l l l l l l l l l} 
        \toprule
        \begin{tabular}{@{}c@{}}\textbf{Accelerator} \\  \end{tabular} 
        & \textbf{Year} & \begin{tabular}{@{}c@{}} \textbf{Platform} \\  \end{tabular} 
        
        & \begin{tabular}{@{}c@{}} \textbf{Workloads} \\  \end{tabular}
        & \textbf{Workload-based H/W Customization}

        & \begin{tabular}{@{}l@{}} \textbf{H/W Reconfiguration} \end{tabular} & \textbf{Scheduling}
        \\ 
        \midrule
        
        f-CNN$^\text{x}$~\cite{fcnnx2018fpl} & 2018 & FPGA 
        & CNNs & Static & \xmark & Static \\
        
        \texttt{PREMA}~\cite{prema2020hpca} & 2020 & ASIC 

        & CNNs/RNNs/Dynamic RNNs & \xmark 
 
        & Preemption/Dynamic & Dynamic \\
        
        AI-MT~\cite{aimt2020isca} & 2020 & ASIC 

        & CNNs/RNNs & \xmark 
        & Preemption/Dynamic & Dynamic \\
        
        \texttt{Planaria}~\cite{planaria2020micro} & 2020 & ASIC 

        & CNNs/RNNs & \xmark 

        & Hardware/Dynamic & Dynamic \\
        
        \texttt{FGSpMt-NPU}~\cite{dataflow_mirroring2021dac} & 2021 & ASIC 

        & CNNs/Transformers & \xmark & Hardware/Dynamic & Dynamic \\
        
        \texttt{Herald}~\cite{hda2021hpca} & 2021 & ASIC 

        & CNNs/RNNs/Segmentors & Static & \xmark & Static \\
        
        \bottomrule
    \end{tabular}
    }
    \label{tab:multidnn-accel-comparison}
\end{table*}

\subsubsection{{Graph NNs}}
{
The recent progress in graph neural networks (GNNs) has led to their broad usage for the processing of graph-organized data.
From a workload perspective, GNNs come with unique challenges: \textit{i)}~the diversity of GNN models, which may include edge, vertex or graph-wide updates and various aggregation functions. This affects both the type (\textit{e.g.}~dense or sparse) and distribution of computation across the graph; \textit{ii)}~the dependence of the workload on the input graph, which affects the size, sparsity and shape of feature vectors. Each GNN's balance of dense and sparse computations and compute- and memory-bound operations poses a challenge in deriving a suitable hardware accelerator; and \textit{iii)}~the diverse performance requirements, from throughput-driven recommender systems with large-scale graphs to low-latency object or fraud detection. Placing throughput or latency first determines the type of optimizations that can be exploited by the underlying hardware. Overall, challenges \textit{i)} and \textit{ii)} impact crucial architectural decisions, such as the allocation of resources for dense or sparse processing elements (PEs), the selection of dataflow and the scheduling policy, while \textit{iii)} sets constraints on the hardware-level techniques that can be employed.
}
\\\textbf{Implications:}
GNNs differ significantly from the well-studied CNN and RNN workloads and call for tailored hardware solutions that address their scalability and performance needs, especially in application with multiple GNNs or with GNNs co-located with other DNNs.


\subsubsection{{NAS-generated DNNs}}
The expanding adoption of neural architecture search (NAS) for the design of highly accurate models can further increase the workload heterogeneity of future DNNs. The automated nature of NAS often leads to nonintuitive topologies with complex and irregular connectivity among layers, up to the extreme of randomly wired networks. 
\\\textbf{Implications:} Compiling these DNNs or deriving even a single-DNN custom accelerator becomes a difficult task. This challenge propagates to the design of multi-DNN accelerators, by further diversifying the workloads that need to be served.

\subsubsection{{Variably Quantized DNNs}}
The broad use of quantization for model compression imposes another dimension of heterogeneity. Different DNNs exhibit variable resilience to quantization, with some models quantized down to very narrow bitwidths (\textit{e.g.}~binary or ternary), while others tolerating only 32-bit or 16-bit floating-point data types without severely degrading the accuracy. 
\\\textbf{Implications:} Multi-DNN accelerators are faced with the major challenge of supporting models quantized in potentially different wordlengths while sustaining high resource utilization and without penalizing performance.

Overall, the rapid algorithmic progress from the ML community requires future-proof solutions that can be re-used by both the already existing diverse models and the future generations of DNNs. Importantly, such hardware solutions need to handle the often-contradictory characteristics that these workloads impose to their accelerators. At the same time, as high performance often requires customization, architects of multi-DNN accelerators are challenged with finding a balance between programmability and customization.

\subsection{\large\textbf{2.2 Performance Criteria}}
Despite having a few \textit{common} performance criteria with single-DNN accelerators, the nature of multi-DNN hardware architectures comes with an additional set of \textit{distinct} metrics. 

\textbf{Common metrics} comprise \textit{i)}~hardware-oriented metrics, including area in mm$^2$ for ASICs or resource consumption for FPGAs affecting form factor and cost, total spent energy in joules (J) that affects battery life or electricity bills, and peak power consumption in watts (W) which has a direct impact on the thermal design and cooling requirements of the system; and \textit{ii)}~user-oriented metrics for single DNNs, such as the quality of service (QoS) experienced by a single user or DNN, expressed as service-level agreement (SLA) violation rate; the SLA typically defines either latency or throughput targets. Moreover, an important user-level metric constitutes the model \textit{accuracy}, which becomes especially relevant when approximate computing techniques are introduced.

\textbf{Distinct metrics} for multi-DNN systems also involve hardware- and user-oriented metrics. Hardware-oriented metrics include the combined \textit{system throughput} (STP) in inferences per second (inf/s) aggregated across all DNNs which indicates the utilization efficiency of the accelerator; and the \textit{execution time improvement} (speedup) per DNN over a baseline that quantifies the benefits of a multi-DNN accelerator over its single-DNN counterpart.
User-oriented metrics include the \textit{normalized turnaround time} (NTT) and its arithmetic \textit{average} (ANTT) that capture the slowdown of a DNN due to the multi-DNN co-location compared to its exclusive execution; and \mbox{\textit{fairness}}, in both its priority-agnostic and priority-aware versions, that assesses how well the resources are balanced. This extended set of metrics play a crucial role in deployability and hence call for new methodologies that optimize them when designing multi-DNN accelerators.

\begin{figure*}[t]
    \centering
    {
    \includegraphics[trim={0cm 1.9cm 0cm 4cm},clip,width=0.9\textwidth]{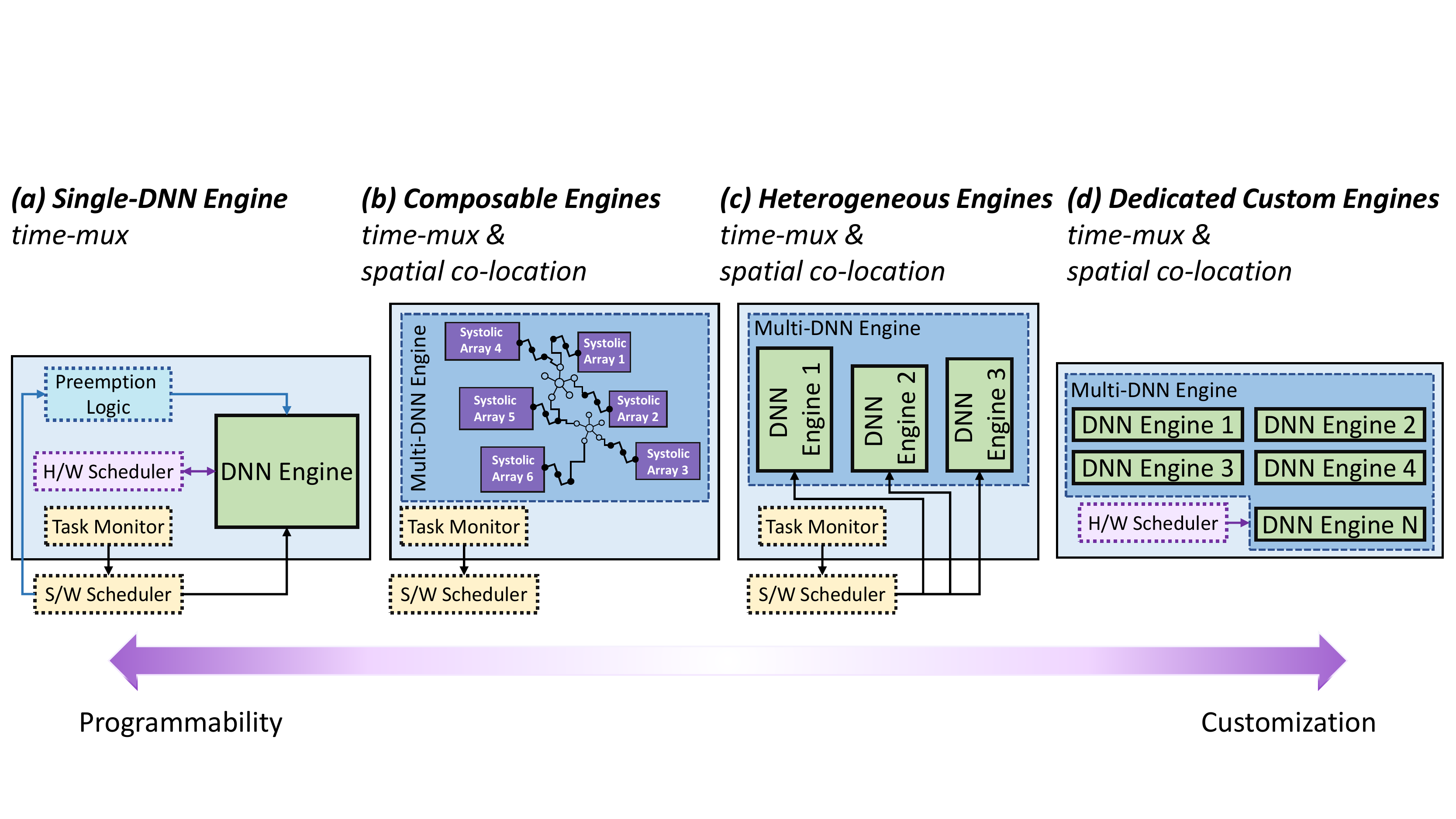}
    }
    \caption{Design space of multi-DNN accelerators.}
    \label{fig:multidnn-hw}
     \vspace{-0.4cm}
\end{figure*}

\section{3 DESIGNING A MULTI-DNN ACCELERATOR}
\label{sec:multi-dnn-hw}

Recently, a handful of works have paved the way towards a new class of multi-DNN hardware acceleration architectures (\textbf{Table}~\ref{tab:multidnn-accel-comparison}).
{Key design decisions} comprise \textit{1)}~the inter-DNN parallelization strategy, \textit{2)}~the design of the DNN engine, and \textit{3)}~the scheduling policy.

\subsection{\large\textbf{3.1 Inter-DNN Parallelization Strategy}}
\label{sec:parallel-strategy}

The parallelization strategy dictates how DNNs are allowed to utilize the resources of the accelerator. We taxonomize the different approaches based on whether they exploit temporal or spatial parallelism, as follows:

\subsubsection{{Time-multiplexing}}
On the one end of the spectrum lie time-multiplexing schemes~\cite{prema2020hpca}. In this case, a \textit{single DNN} occupies the full resources of the accelerator (\textbf{Figure}~\ref{fig:multidnn-hw}a), consisting of both the PEs and the off-chip memory bandwidth. This approach confines the optimization to the scheduling level, with systems such as \texttt{PREMA}~\cite{prema2020hpca} leveraging preemptive policies with execution time prediction to yield optimized multi-DNN schedules.

An enhancement over strict exclusive access comprises \textit{inter-DNN pipelining}~\cite{aimt2020isca}, allowing the occupancy of the PEs by one DNN and the memory bandwidth by another. Systems such as AI-MT~\cite{aimt2020isca} overlap the computation of compute-bound layers from one DNN with the communication of memory-bound layers from another. This approach aims to eliminate resource idleness by maximizing the sustained utilization of both PEs and off-chip memory bandwidth. 

Overall, time-multiplexing strategies focus on optimizing the scheduling and require minimal to no changes at the hardware level to support the execution of multiple DNNs. As such, they need to be used in cases where existing accelerators are to be used or hardware modifications are costly.

\subsubsection{{Spatial Co-location}}
An orthogonal approach to inter-DNN time-multiplexing is their spatial co-location. Under this approach, multiple DNNs occupy different parts of the accelerator and their execution progresses in parallel. 
So far, two main design paradigms have been proposed: \textit{i)}~\textit{dynamically composable engines} and \textit{ii)}~\textit{statically instantiated heterogeneous engines}.

The first line of work has focused on designing reconfigurable architectures that allow the dynamic composition of engines to build larger processing blocks (Figure~\ref{fig:multidnn-hw}b). This technique can be applied either by combining coarse systolic arrays with a \textit{uniform} shape~\cite{planaria2020micro} or at a finer-grained manner using \textit{varying-sized} systolic arrays~\cite{dataflow_mirroring2021dac}. In both of these cases, \textit{resource partitioning} decisions are made \textit{at run time} based on the DNNs that are co-located at any time instant.

Another stream of work has proposed the static instantiation of heterogeneous engines.
Key behind this family of approaches is the assumption that the target set of DNNs is known \textit{a priori} and hence this information can be exploited to further customize the multi-DNN accelerator \textit{at design time}.
\mbox{Herald}~\cite{hda2021hpca} capitalizes upon the fact that different layers map more optimally on designs that exploit different dataflows (\textit{e.g.}~weights vs output stationary) and places up to three \textit{predefined sub-accelerators} with \textit{different dataflows} on the same chip (Figure~\ref{fig:multidnn-hw}c). The sub-accelerators are allocated computational and bandwidth resources based on the workloads of the target DNNs and each layer is dispatched to the most appropriate sub-accelerator as determined by the system's scheduler.
Overall, this methodology leverages pre-existing accelerator designs that have already-demonstrated merits by treating them as templates. Nevertheless, it restricts the level of hardware customization that can be applied, leading to an end system that has the same characteristics as its sub-accelerators.

Towards higher customizability, \mbox{f-CNN$^\text{x}$~\cite{fcnnx2018fpl}} proposes the instantiation of \textit{one custom engine per DNN} (Figure~\ref{fig:multidnn-hw}d). Each engine is tailored to the workload of its DNN and the resource partitioning between engines is optimized at design time. 
This approach places fairness at the forefront, as all DNNs are executed in parallel and the resource partitioning follows the relative performance requirements of the DNNs, and allows for a finer granularity of control at both the microarchitectural and resource allocation levels.

Compared to solely using time-multiplexing, spatial co-location approaches are more invasive as they require a complete change of the underlying hardware design. Nonetheless, the hardware modifications often come with increased performance and efficiency~\cite{fcnnx2018fpl,planaria2020micro,dataflow_mirroring2021dac,hda2021hpca}.

\subsection{\large\textbf{3.2 DNN Engine Microarchitecture}}
\label{sec:microarch}

Two main DNN engine paradigms have emerged that aim at different objectives. The first paradigm aims to efficiently make \textit{existing} DNN accelerators support multiple DNNs. The second focuses on designing \textit{custom} accelerators, explicitly optimized for multi-DNN execution.

\subsubsection{{Enhancing Existing Architectures}}
With the wider availability of DNN accelerators, a few works have proposed enhancements that lightly modify the existing single-DNN hardware architectures to enable multi-DNN execution. Typically, the core DNN engine comprises a single processing engine that adopts a fixed dataflow and is time-shared between (sub-)layers. 
\texttt{PREMA}~\cite{prema2020hpca} introduces a preemption module that resides next to the core DNN engine (\textit{Preemption Logic} in Figure~\ref{fig:multidnn-hw}a) and is responsible for efficiently enforcing a preemptive scheduling policy. The scheduling decisions are made from a software runtime, which in turn configures the hardware preemption module.
Bringing more functionality into hardware, AI-MT~\cite{aimt2020isca} introduces a hardware scheduler that constantly monitors the state of the DNN engine and makes scheduling decisions to coordinate the execution (\textit{H/W Scheduler} in Figure~\ref{fig:multidnn-hw}a). The scheduler consists of a state machine that implements the AI-MT's scheduling algorithm, together with the supporting data structures. 
As such, after an initial configuration upon the system startup, AI-MT's multi-DNN scheduling takes place fully in hardware without the mediation of software.

\subsubsection{{Devising Custom Architectures}}
When designing custom architectures for multiple DNNs, the spatial co-location strategy dominates as it opens an additional optimization dimension.
\textbf{\mbox{Dynamically} composable engines} introduce omnidirectional connections between the PEs of each systolic array (Figure~\ref{fig:multidnn-hw}b). This is implemented by means of lightweight switches that select the inputs and direct the outputs of each PE.
\texttt{Planaria}~\cite{planaria2020micro} considers uniformly sized systolic arrays and groups them into multiple pods. All arrays within a pod share the same on-chip memory and are connected via a crossbar, while the pods communicate through a bi-directional ring bus.
This hierarchical organization enables the efficient cooperation between same-pod arrays and does not penalize the clock frequency.

Despite its merits, the arrays' uniform size restricts the customization potential and can lead to underutilization.
To counteract this, FGSpMt-NPU~\cite{dataflow_mirroring2021dac} proposed the dataflow-mirroring technique. Instead of composing fixed-sized arrays into larger structures, this scheme combines individual rows and columns of the arrays to build arbitrarily sized blocks. To enable this functionality, the PEs are equipped with additional switches towards all four adjacent PEs.

\textbf{Statically selected and coarse heterogeneous engines} rely on the utilization of pre-existing accelerator designs (Fig.~\ref{fig:multidnn-hw}c). Exemplified by Herald~\cite{hda2021hpca}, this design treats pre-existing accelerators as building blocks with different dataflows that operate on different DNNs in parallel and are connected to a shared on-chip buffer through a network-on-chip (NoC). This approach  focuses on the allocation of resources and bandwidth between the selected accelerators, without the need for new hardware modules.

Further towards customizability, \textbf{statically selected and highly customized heterogeneous engines} are derived fully based on the workload characteristics of the target set of DNNs and the per-DNN performance requirements (Figure~\ref{fig:multidnn-hw}d). As manifested in f-CNN$^\text{x}$~\cite{fcnnx2018fpl}, the DNN engines follow a streaming design, \textit{i.e.}~instead of using a single processing engine that is time-shared across layers, each engine can have an arbitrary pipeline of coarse stages with DNN layers being pipelined. The stages of the pipeline, the connectivity with each other and the resource allocation across stages can be customized to the given DNN. To coordinate the operation of the multiple DNN engines, f-CNN$^\text{x}$ introduces a hardware scheduler (\textit{H/W Scheduler} in Figure~\ref{fig:multidnn-hw}d) that deterministically allocates off-chip memory bandwidth to the engines. Overall, the design of each engine, the resource partitioning across them and the memory bandwidth allocation are co-optimized statically at design time, leading to a full-custom accelerator for a given set of DNNs.

\subsection{\large\textbf{3.3 Scheduling Algorithms}}
\label{sec:scheduler}

The scheduling algorithms can be classified along three dimensions: \textit{i)}~whether they happen statically at design time or dynamically at run time, \textit{ii)}~whether they control the temporal or spatial mapping and the resource partitioning, and \textit{iii)}~whether they are hardware- or software-based.

\subsubsection{{Static vs. Dynamic Scheduling}}
Depending on the high-level objectives of the use-case, static or dynamic scheduling is more appropriate.
Systems that prioritize flexibility tend to adopt dynamic scheduling. This approach allows new DNN inference tasks to be served and resources to be re-allocated based on the completed and currently running DNNs. As a result, accelerators that rely on time-multiplexing~\cite{prema2020hpca,aimt2020isca} or employ dynamically composable engines~\cite{planaria2020micro,dataflow_mirroring2021dac} utilize dynamic schedulers, running either in software~\cite{prema2020hpca,planaria2020micro,dataflow_mirroring2021dac} or hardware~\cite{aimt2020isca}.

On the other hand, hardware designs that favor customization to deliver high performance adopt static scheduling. In this case, the set of target DNNs has to be known \textit{at design time} in order to generate an optimized execution schedule. At run time, the static schedule is implemented through the control of software~\cite{hda2021hpca} or a dedicated hardware module~\cite{fcnnx2018fpl}.
The merits of this approach is that more optimization opportunities can be exploited, \textit{e.g.}~some layers can be suboptimally scheduled in order to obtain higher performance globally~\cite{hda2021hpca}, and actual performance is more predictable due to the deterministic schedule and the lack of cross-DNN interference. 
Additionally, the fixed nature of the schedules makes this approach suitable for fixed-purpose systems or when workloads change across long time scales, such as autonomous vehicles and mobile robots.

\subsubsection{{Scheduling Decisions}}
Different systems place a varying level of complexity to their scheduler.
Accelerators that do not modify the core DNN engine tend to employ more sophisticated algorithms, such as \texttt{PREMA}'s preemptive policy~\cite{prema2020hpca} and AI-MT's early-eviction method~\cite{aimt2020isca}. In this setting, the scheduler determines which (sub-)layer of which DNN is to run next on the fixed DNN engine, and optionally can launch the concurrent data transfer for the (sub-)layer of another DNN to utilize the off-chip memory bandwidth more fully.

In contrast, systems that focus more on spatial co-location and optimizing the processing engine tend to adopt \textit{heuristic} algorithms which determine both the temporal and the spatial mapping of DNNs~\cite{planaria2020micro,dataflow_mirroring2021dac,hda2021hpca}. For instance, the schedulers of \texttt{Planaria}~\cite{planaria2020micro} and FGSpMt-NPU~\cite{dataflow_mirroring2021dac} dynamically determine the composition of engines, effectively re-allocating the available resources among DNNs when an inference task finishes or a new one arrives.
Alternatively, the scheduler in~\cite{hda2021hpca} jointly determines a static layer order of execution and task-to-engine mapping to orchestrate the processing of its diverse DNN engines.
An exception to heuristic approaches is f-CNN$^\text{x}$~\cite{fcnnx2018fpl} which imposes deterministic, hardware-controlled policies based on a cyclic scheduling formulation. However, this approach is feasible in use-cases where the DNNs and their performance requirements are fixed and known before deployment.

\section{4 THE ROAD AHEAD}
\label{sec:challenges-future}

\vspace{0.2cm}
\subsubsection{Customization vs. Programmability}
\label{sec:perf-vs-flex}

Peak performance is often reached through fine-grained customizability~\cite{fcnnx2018fpl,hda2021hpca} at the cost of a new design cycle when a different set of DNNs is targeted. This approach is feasible under three settings: \textit{i)}~hardware reconfigurability, \textit{ii)}~design automation, and \textit{iii)}~slow rate of change of application. 

When reconfigurable platforms, such as \mbox{FPGAs}, are targeted, the fabric can be reprogrammed with a different design within a few 100s of milliseconds in the occurrence of a new set of DNNs. This can be observed in the FPGA-based flow of f-CNN$^\text{x}$ and constitutes a key difference to ASIC-based solutions where a new chip must be fabricated to adopt a new hardware design.
Similarly, deriving a new hardware design for each set of DNNs requires scalable multiple DNNs-to-hardware design flows that \textit{automate} the accelerator generation process~\cite{fcnnx2018fpl,hda2021hpca} and remove the need for excessive engineering hours. Finally, a slow rate of change in applications, \textit{e.g.}~the set of DNNs and their performance requirements, is often required to justify the customization with respect to specific DNNs. This setting is currently present in \textit{single-purpose embedded systems}, such as mobile robots, where the multiple concurrent tasks largely remain constant in the long term, but does not apply to \textit{cloud-based systems}, where the DNN workloads that need to be served evolve continuously.

On the other end, ASIC-based solutions require future-proof designs that can amortize the high upfront fabrication cost through re-use of the chip across broad DNN workloads. To enable this, existing approaches either co-schedule diverse DNNs with complementary workload properties~\cite{prema2020hpca,aimt2020isca} (\textit{e.g.}~memory-bound RNNs and compute-bound CNNs) or introduce \textit{soft} hardware reconfiguration, through software-programmable switches~\cite{planaria2020micro,dataflow_mirroring2021dac} that do not require hardware changes.

The current design flows adopt a strict stance with respect to workload-based customization (Table~1); they either re-design the accelerator~\cite{fcnnx2018fpl,hda2021hpca} or they utilize a fixed hardware architecture for all cases~\cite{prema2020hpca,aimt2020isca,planaria2020micro,dataflow_mirroring2021dac}. As such, new methodologies that balance \textit{hard design-time customization} and \textit{soft run-time programmability} based on the target use-case can provide a more universal design approach that covers a more comprehensive design space. With the increased design space comes aggravated optimization complexity and hence research is required into developing efficient methodologies that yield high-performance multi-DNN accelerators.

\subsubsection{Approximate Computing for Multiple DNNs}
\label{sec:approx-computing}

Another direction for enhancing the performance and energy efficiency of multi-DNN accelerators is through approximation-based techniques~\cite{approx_hw2019csur}. Such schemes, equivalently interpreted as a form of model compression, include techniques, such as low-precision quantization, data-~\cite{def2021aspdac} and weights-reduction~\cite{unzipfpga2021fccm} methods, that would be applied in a \textit{non-uniform manner} across the DNNs based on the degree of approximation tolerated by each DNN before degrading its accuracy below a user-defined acceptable level. 

For multiple DNNs, we envisage the new paradigm of \textit{cross-DNN approximate computing} which entails methods that exploit \textit{cross-DNN redundancy}, \textit{workload commonalities} or \textit{differences in resilience to approximation} across the models. Although a few works have looked into such methods in the multi-DNN context, including dynamically scalable DNNs~\cite{nestdnn2018mobicom} and cross-DNN weights sharing~\cite{virt_weights2020mobisys}, these are tailored for general-purpose processors and do not allow for hardware-level optimizations.

An early approach in this direction was presented in~\cite{multilstms2020fpt} in the context of multi-LSTM applications. The proposed scheme tunably decomposes the weight matrices of multiple LSTM models and represents them with a shared low-rank representation. The degree of decomposition is jointly optimized with the underlying accelerator design, to yield a multi-LSTM hardware architecture that fully capitalizes over the theoretical gains of the induced approximations.

In general, general-purpose processors struggle to materialize the benefits of approximate computing approaches. This is often the case due to the workload irregularity caused by approximations (\textit{e.g.}~fine-grained DNN pruning) or the need for specialized processing units (\textit{e.g.}~low-precision data types). Custom-built multi-DNN accelerators can play a key role in enabling the development of new cross-DNN approximate algorithms that lead to controlled accuracy drop and high processing speed.

\subsubsection{Multi-DNN Model-Hardware Co-design}
Towards extracting both maximum performance and accuracy, model-hardware co-design approaches have begun to gain traction. 
Co-design allows us to exploit shared trade-offs between model and hardware in order to develop higher-performing end-to-end systems. Despite the increased range of optimization opportunities, this approach also comes with a large design space that is not trivial to navigate, hence calling for principled and scalable solutions.

In the context of multiple DNNs, co-design methodologies need to consider the multiple AI tasks of an overarching application and design from scratch both the DNN models and the multi-DNN accelerator. An early work in this direction is ASICNAS~\cite{asicnas2020dac}. To tackle the exponential design space, ASICNAS treats a set of pre-defined hardware architectures as templates, and considers only these in its search. ASICNAS' approach demonstrates 2$\times$ energy reduction with less than 1.6 percentage points accuracy drop, showcasing the potential of co-design schemes in pushing further the performance of multi-DNN systems.

Nevertheless, the primary challenge that obstructs co-design methods is still standing: the excessively high-dimensional design space that encompasses model-, scheduling- and hardware-level parameters, which is further aggravated by the multiplicity and variability of DNNs. As such, further research effort needs to be invested in developing scalable methodologies that overcome this complexity in order to lead to the next-generation of multi-DNN platforms.

\section{5 CONCLUSION}
\label{sec:conclusion}

As multi-DNN AI applications are rapidly becoming popular, existing single-DNN accelerators fail to provide the required performance. As such, there is an emerging need for a paradigm shift towards multi-DNN accelerator design. Primary challenges in this endeavor constitute the workload diversity of future DNNs and the new set of multi-DNN performance metrics. At the hardware front, the key design decisions differ from those of single-DNN accelerators, with the inter-DNN parallelization approach and the scheduling policy coming at the forefront. To ensure the usability of multi-DNN accelerators, we argue that balancing customization with flexibility and developing the necessary software support are among the main pillars.
Finally, we highlight how approximate computing techniques that exploit cross-DNN commonalities and model-hardware co-design methodologies that can scale to multiple DNNs are key drivers towards performant and efficient multi-DNN accelerators.

\bibliographystyle{IEEEtran}
\bibliography{references.bib}

\begin{IEEEbiography}{Stylianos I. Venieris}{\,} is currently a Senior \mbox{Researcher} at Samsung AI, Cambridge, UK. He received the Ph.D. degree from Imperial College London, U.K. He is a Member of IEEE and ACM. Contact him at s.venieris@samsung.com.
\end{IEEEbiography}

\begin{IEEEbiography}{Christos-Savvas Bouganis}{\,} is currently a Reader in Intelligent Digital Systems with Electrical and Electronic Engineering Department, Imperial College London, London, U.K. He is Senior Member of IEEE. Contact him at ccb98@ic.ac.uk.
\end{IEEEbiography}

\begin{IEEEbiography}{Nicholas D. Lane}{\,} is an Associate Professor in the Department of Computer Science and Technology, University of Cambridge, U.K., and a Program Director at the Samsung AI Center, Cambridge, U.K. Contact him at ndl32@cam.ac.uk.
\end{IEEEbiography}

\end{document}